\documentstyle[12pt]{article}
\newcommand{\tcite}[2]{\cite{#1, #2}}
\newcommand{\mlabel}[1]{\label{#1}}
\newcommand{\mibitem}[1]{\bibitem{#1}}

\newcommand{\be}{\begin{equation}}
\newcommand{\ee}{\end{equation}}
\newcommand{\ba}{\begin{eqnarray}}
\newcommand{\ea}{\end{eqnarray}}
\newcommand{\bastar}{\begin{eqnarray*}}
\newcommand{\eastar}{\end{eqnarray*}}
\newcommand{\half}{{1 \over 2}}
\newcommand{\del}{\partial}
\newcommand{\tensor}{\otimes}
\newcommand{\f}[1]{f^{#1}}
\newcommand{\poisson}[2]{\{ #1 , #2 \}}
\newcommand{\commut}[2]{\lbrack #1 , #2 \rbrack}
\newcommand{\set}[1]{\{#1\}}
\newcommand{\bartheta}{{\bar\theta}}
\newcommand{\barpi}{\bar\pi}
\newcommand{\barphi}{\bar\phi}
\newcommand{\bareta}{\bar\eta}
\newcommand{\g}{{\bf g}}
\newcommand{\A}{{\cal A}}
\newcommand{\C}{{\cal C}}
\newcommand{\E}{{\cal E}}
\newcommand{\X}{{\cal X}}
\newcommand{\p}{{\cal P}}
\newcommand{\D}{{\cal D}}
\newcommand{\G}{{\cal G}}
\newcommand{\barp}{\bar{\cal P}}
\newcommand{\cL}{{\cal L}}
\newcommand{\w}{{\cal W}}
\newcommand{\h}{{\cal H}}
\renewcommand{\.}{\ .}
\renewcommand{\,}{\ ,}
\newcommand{\refe}[1]{(\ref{#1})}

\newcommand{\vali}{\vspace{7mm}}

\newcommand{\ie}{{\it i.e.\ }}
\newcommand{\wsigma}{\cite{wsigma}}

\newcommand{\dh}{\cite{dh}}

\newcommand{\ab}{\cite{ab}}
\newcommand{\osvb}{\cite{osvb}}
\newcommand{\kalkman}{\cite{kalkman}}

\newcommand{\kanno}{\cite{kanno}}

\newcommand{\bfv}{\cite{bfv}}
\newcommand{\hnt}{\cite{hnt}}

\begin{document}
\begin{titlepage}
\begin{flushright}
UU-ITP 01/94 \\
HU-TFT-93-65 \\
hep-th/9403126
\end{flushright}

\vskip 0.7truecm

\begin{center}
{ \bf EQUIVARIANCE, BRST AND SUPERSPACE \\  }
\end{center}

\vskip 1.5cm

\begin{center}
{\bf Antti J. Niemi $^{*}$ } \\
\vskip 0.4cm
{\it Department of Theoretical Physics, Uppsala University
\\ P.O. Box 803, S-75108, Uppsala, Sweden }
\vskip 0.4cm
and \\
\vskip 0.4cm
{\bf Olav Tirkkonen $^{**}$ } \\
\vskip 0.4cm
{\it Research Institute for Theoretical Physics \\
P.O. Box 9, FIN-00014 University of Helsinki, Finland}\vskip 0.2cm
\end{center}

\vskip 2.7cm

\rm
\noindent
The structure of equivariant cohomology in non-abelian localization
formulas and topological field theories is discussed. Equivariance is
formulated in terms of a nilpotent BRST symmetry, and another
nilpotent operator which restricts the BRST cohomology onto the
equivariant, or basic sector.  A superfield formulation is presented
and connections to reducible (BFV) quantization of topological
Yang-Mills theory are discussed.

\vfill

\begin{flushleft}
\rule{5.1 in}{.007 in}\\
$^{*}$ {\small E-mail: NIEMI@RHEA.TEORFYS.UU.SE \\}
$^{**}$ {\small E-mail: OLAV.TIRKKONEN@HELSINKI.FI \\ }
\end{flushleft}

\end{titlepage}

\vfill\eject

\baselineskip 0.65cm

\noindent
{\bf 1. Introduction }

\vskip 1.0cm

In the present paper we study equivariant ({\it i.e.} basic)
cohomology, a concept which is acquiring increasing
attention in modern theoretical physics. This interest is
mainly a consequence of the relevance that equivariant
cohomology has to various localization formulas, quantum
integrability and topological field theories.

Here we shall be mostly interested in equivariant cohomology
relevant to localization formulas. However, we shall also
investigate relations to topological field theories, and in
particular four dimensional topological Yang-Mills theory.

The original localization formula by Duistermaat and Heckman
(DH) \dh \ con\-cerns exponential integrals over symplectic
manifolds, \ie classical partition functions.  The theorem
states that if a Hamiltonian $H$ on a $2n$-dimensional
compact symplectic manifold with symplectic two-form
$\omega$ generates the Poisson action of a torus, the
stationary phase approximation is exact
\[
\int \omega^{n} \exp\{ -i\phi H \} \ = \
\frac{1}{\phi^n}\sum\limits_{dH=0} { \exp(-i\phi H ) \over
\sqrt{ det|| \partial_{ij}  H || } }
\]
In {}\tcite{bv}{ab} it was noted that the underlying structure
in the DH formula was that of equivariant cohomology with
respect to the torus acting on the manifold: The integrand
depends only on the equivariant extension $ \omega + \phi
H$ of the symplectic two-form, and the exactness of
stationary phase approximation is a consequence of an
equivariant version of Stokes theorem.

Infinite dimensional generalizations of DH were introduced in
\tcite{atiyah}{bismut} and loop space extensions were considered in
\cite{omat}.  In particular, in these papers localization proofs of
index theorems for Dirac operators and their equivariant extensions
with respect to Lie groups were related to equivariant cohomology.
The
formalism was also shown to be relevant in a geometric formulation of
Poincare supersymmetric field theories {}\cite{palo}.

A generalization to include a non-abelian group action was
presented by Witten in {}\cite{wtwodim}, who applied it to
two dimensional Yang-Mills theory. In {}\cite{onexact} this
approach was combined with {}\cite{omat}, to localize path
integrals with Hamiltonians that are {\it a priori}
arbitrary functions of generators of circle actions. In
particular, the equivariant exterior derivative was written
as a sum of a nilpotent part related to the group action,
and a kinetic part related to a model independent loop space
circle action.

\vali

For a long time, equivariant cohomology has also played a
central r\^ole in topological field theories \cite{bbrt}.
Indeed, topological field theories share many
aspects with the generalized Duistermaat-Heckman systems;
Most notably their path integrals can be evaluated exactly
using a localization method. Topological field theories
also possess a nilpotent BRST-like symmetry, and they can be
viewed as BRST gauge fixings of underlying trivial theories.

In \wsigma \ it was first realized that the structure determined by
the
(nilpotent) BRST-operator in four dimensional topological Yang-Mills
theory
{\cite{wtym}} is that of equivariant cohomology.  The argument was
strenghtened
in \osvb, where the BRST structure was identified with basic
cohomology, and in
\kanno \ where the Weil algebra structure of topological Yang-Mills
theory was
clarified.

The approach of \osvb \ resolves elegantly the problem of
nontriviality of observables in topological Yang-Mills
theory.  In particular, the structure of basic cohomology
introduced there shows that in addition of computing
the BRST cohomology it is necessary to restrict onto the
basic forms.  In \kalkman \ the approach of \osvb \ was
subsequently related to the mathematical theory of BRST
\cite{koster}.

\vali

In the following we develop the BRST description of equivariant
cohomology from the point of view of localization formulas. We are
particularly interested in exposing the intimate relationship between
the equivariant stucture underlying localization formulas and
topological Yang-Mills theory, as formulated in {}\tcite{mn}{hnt}.
For
this, we first briefly review symplectic action of a Lie group in
Section 2.  In section 3. we present the different models of
equivariant cohomology and in section 4. we consider the construction
of equivariant cohomology operators and in particular equivariant
extensions of the symplectic two-form in the framework of
localization formulas. In Section 5. we introduce a superfield
formalism and in
section 6. we discuss the connection of the construction with the
Batalin-Fradkin-Vilkovisky approach to reducible constrained systems.

\vfill\eject
\noindent
{\bf 2.  Symplectic actions and Lie groups}

\vskip 1.0cm

In the present paper we are interested in the equivariant
cohomology which is related to the action of a connected Lie
group $G$ as local diffeomorphisms on a symplectic manifold
$M$. The dimension of $M$ is $2n$, and local coordinates on
$M$ are denoted by $z^k, k=1 \ldots 2n$.

The $G$-action on $M$ is generated by vector fields $\X_a, a = 1
\ldots m$ that realize the commutation relations
of $G$,
\be
\commut{\X_a}{\X_b} ~=~ f^{abc} \X_c \
\mlabel{vcomm}
\ee
with $f^{abc}$ the structure constants of the Lie algebra $\bf g$ of
$G$.

With $\X$ a generic vector field on $M$, we denote contraction along
$\X$ by $i_\X$. In particular, the basis of contractions
corresponding
to the Lie algebra generators $\lbrace \X_a \rbrace$ is denoted by
$i_{\X_a}
\equiv i_a$.  The pertinent Lie-derivatives
\[
\cL_{a}  = d i_{a} + i_{a} d
\]
with $d$ the exterior derivative on the exterior algebra $\Omega(M)$
of $M$, then generate the $G$-action on $\Omega(M)$,
$$
\commut{\cL_a}{\cL_b} = f^{abc} \cL_c \  .
$$
In addition we have the Lie-derivative action on the contraction:
\[
\commut{i_a}{\cL_b} = f^{abc} \ i_c \  .
\]

The symplectic two-form
\[
\omega = \half\omega_{kl} dz^k \wedge dz^l
\]
on $M$ is closed and nondegenerate. Locally,
$$
\omega = d \vartheta \  .
$$
where the one-form $\vartheta$ is  the symplectic potential.
We shall assume that the action of $G$ is symplectic so that it
preserves the symplectic structure,
\be
\cL_a \omega ~=~ d i_a \omega ~=~ 0 \  . \mlabel{symplact}
\ee
If the one-forms $i_a  \omega$ are exact
(for this the triviality of  $H^1(M,R)$ is sufficient), we can
then introduce the momentum map {}\cite{gs} $ H : M \mapsto {\bf
g}^{*}$
where ${\bf g}^{*}$ is the dual Lie algebra. When evaluated on a
vector
field $\X$, the momentum map $H$
yields the corresponding Hamiltonian $H_{\X}(z)$ by
\be
i_\X \omega = -d H_\X
\mlabel{hamvfield}
\ee
or in local coordinates,
$$
\X= \omega^{kl} \partial_k H_\X \del_l \  .
$$ For the Lie algebra $\bf g$ this yields a one-to-one
correspondence
between the vector fields $\X_a$ (and corresponding Lie-derivatives
$\cL_a$) and certain functions $H_a$ on $M$, the components of the
momentum map
\be
H = \phi^a H_a \, \mlabel{mommap}
\ee
where $\{ \phi^a \}$ is a (symmetric) basis of the dual Lie algebra
${\bf g}^*$.

The Poisson bracket of the Hamiltonians $H_a$ is defined by
$$
\poisson{H_a}{H_b}  = \omega(\X_a, \X_b) =\cL_a
H_b.
$$
{}From the Jacobi identity for $\bf g$ we then get the
homomorphism
$$
\X_{\poisson{H_a}{H_b}} = \commut{\X_a}{\X_b} \  .
$$
However, the inverse is not necessarily true: The Hamiltonian
function which corresponds to the commutator of two group generators
may differ from the Poisson bracket of the pertinent Hamiltonian
functions,
\be
\poisson{H_a}{H_b} = \f{abc} H_c + \kappa_{ab} \  .
\mlabel{cocycle}
\ee
Here $\kappa_{ab}$ is the 2-cocycle in the Lie-algebra cohomology of
$\bf g$, and the appearance of a cocycle can be related to the
possible
noninvariance of the symplectic potential under $G$:  From
(\ref{symplact}) it follows that
\be
i_{a} \vartheta = H_a + h_a \mlabel{smallh}
\ee
with some functions $h_a$ on $M$, and the definition of the Poisson
bracket
implies the cocycle in (\ref{cocycle}) is given by
\be
\kappa_{ab} = \f{abc} h_c - \cL_a h_b + \cL_b h_a
\mlabel{kappeq}
\ee
Thus, only when $\kappa_{ab} = 0$ for all $a,b$ does the
$G$-action
of the vector fields $\X_a$ lift  isomorphically to the Poisson
action of the
corresponding Hamiltonians $H_a$ on $M$.

\vskip 1.0cm

\noindent
{\bf 3. Models for equivariant cohomology}

\vali

We are interested in the equivariant cohomology $H^*_G(M)$ associated
with the symplectic action of the Lie group $G$ on the manifold
$M$. For this we first note that $H^*_G(M)$ is essentially the deRham
cohomology of $M ~mod(G)$.  If the action of $G$ is free {\it i.e.}
the only element of $G$ which acts trivially is the unit element, the
quotient space $M/G$ is well defined and the $G$-equivariant
cohomology of $M$ coincides with the ordinary cohomology of $M ~ mod
(G)$, that is $H^*_G(M) = H^*(M/G)$.

For the non-free action of a compact group $G$ there exists three
different approaches to model $H^*_G(M)$ using differential forms on
$M$ and polynomial functions and forms on the Lie algebra $\g$ of
$G$.
The two classical models are the Cartan and Weil ones, described {\it
e.g.} in {}\tcite{ab}{mq}.  These two are interpolated by the BRST
model, which is relevant to the BRST structure of topological field
theories.  The BRST model is discussed {\it e.g.}  in \osvb \ and
\kalkman \ and the interrelations between the different models are
clarified in \kalkman.

\vskip 0.4cm
{\it The Cartan Model:} The simplest example of a group
action on the symplectic manifold $M$ is that of the action
of the circle $ G= S^1 = U(1) $, determined by a vector
field $\X$ as the generator of the Lie-algebra $u(1)$ of
$U(1)$. In order to describe the corresponding equivariant
cohomology of $M$ we introduce the following equivariant
exterior derivative operator on $M$
{\tcite{bv}{wdiffg}} \
\be
s = d - \phi \ i_\X \  . \mlabel{sabelcartan}
\ee
where the sign is chosen for later convenience.  The factor
$\phi$ is a real parameter, and the operator $s$ acts on the
whole deRham complex $\Omega(M)$ of differential forms on
$M$.

The square of $s$ is the Lie-derivative with respect to $\X$,
\be
s^2 = -\phi (d i_\X +  i_\X d) = -\phi \cL_\X \ . \mlabel{ssquare}
\ee
Thus on the subcomplex $\Omega_{U(1)}$ of $U(1)$-invariant
exterior forms, $s$ is nilpotent and defines an exterior differential
operator.  The cohomology of $s$ on this subcomplex defines the
equivariant cohomology $H^*_{U(1)}$ of the manifold $M$.

We are especially interested in the equivariant
extension of the symplectic two-form $\omega$. With $H$  the momentum
map of the circle action on $M$, {\it i.e.}  the Hamiltonian
corresponding to $\X$ we get from the definition \refe{hamvfield},
\be
(d - \phi \ i_\X) (\omega - \phi H ) = 0 \ ,
\mlabel{equivham}
\ee
which identifies $\omega - \phi H$ as the equivariant extension of
the
symplectic two-form. Of particular interest are related integrals
over
$M$ that can be evaluated by localization methods based on
{}\refe{equivham}, such as the DH integration formula
\[
\int \omega^n \exp\{ - i\phi H \} ~=~ (-i)^n n! \int \exp\{i(\omega -
\phi H)\}
\]
\be
= ~ \frac{1}{\phi^n}\sum\limits_{dH=0} { \exp(-i\phi H ) \over
\sqrt{ det|| \partial_{ij}  H || } }
\mlabel{dh}
\ee
{}From {\refe{equivham}} we see that the integrand is an
equivariantly
closed form, and localization to the critical points of $H$ follows
from
changing the representative of \refe{equivham} in the equivariant
cohomology
class.

\vskip 0.3cm

In order to generalize for a non-abelian group $G$, the
parameter $\phi$ must first be properly interpreted. For
this we identify it as a generator of the algebra of
polynomials on $u(1)$, {\it i.e.} as a basis element of the
symmetric algebra $S(u(1)^*)$ over the dual of the
Lie-algebra of $U(1)$.  The operator {}\refe{sabelcartan}
then acts on the complex $S(u(1)^*) \tensor \Omega(M)$, and
from (\ref{ssquare}) we conclude that on the
$U(1)$-invariant subcomplex $(S(u(1)^*) \tensor
\Omega(M))^{U(1)}$ \ the action of $s$ is nilpotent, and the
equivariant cohomology is the $s$-cohomology of $(S(u(1)^*)
\tensor \Omega(M))^{U(1)}$.  As shown in {}\ab, the
operations of evaluating $\phi$ and formation of cohomology
commute for abelian group actions, so that the results
coincide independently of the interpretation of $\phi$.
This model for equivariant cohomology is called the abelian
{\it Cartan model}.

For a free  $U(1)$-action we have
\[
(S(u(1)^*) \tensor \Omega(M))^{U(1)}
= S(u(1)^*) \tensor \Omega(M \ mod\ U(1)) \.
\]
{}From this we see that the multipliers $\phi$ play in this case no
cohomological role, and the equivariant cohomology really restricts
to
the cohomology of the quotient space $M\ mod\ U(1)$.

\vali

In order to generalize the Cartan model to non-abelian compact Lie
groups $G$, we consider the algebra $S(\g^*)$ of polynomials on $\g$
as the symmetric algebra on the dual $\g^*$ of $\g$. With $m$ the
dimension of $\g$, the basis of $S(\g^*)$ which is dual to some basis
$\lbrace X_a \rbrace$ of $\g$ is $\phi^a {}~(a = 1, \ldots m)$.

In analogy with (\ref{sabelcartan}) we introduce the
nonabelian equivariant exterior derivative
\be
s = d - \phi^a i_{a} \mlabel{cartandiff}
\ee
which squares to the Lie-derivative
\be
s^2 = - \phi^a \cL_a \,  \mlabel{squs}
\ee
As a consequence $s$ determines a nilpotent exterior derivative
operator on the
$G$-invariant subcomplex
\be
\Omega_G(M) = (S(\g^*) \tensor \Omega(M))^G \  \mlabel{cartancplex}
\ee
of the exterior algebra on $M$. Elements of $\Omega_G(M)$
are equivariant differential forms, that is mappings $\mu$
from $\g$ to $\Omega(M)$ which fulfill the equivariance
condition $\mu(\gamma \cdot X) = \gamma \cdot \mu(X)$ for
$\gamma\in G$, $X \in \g$ and $G$ operating by the adjoint
action on $\g$.

The cohomology of $s$ which we denote $H_s^*(\Omega_G(M))$, then
gives
the Cartan model for the $G$-equivariant cohomology on $M$.

\vskip 0.4cm

{\it The Weil Algebra:}
In order to formulate the equivariant cohomology
using nilpotent operators, we also need to introduce anticommuting
ghosts corresponding to the factors $\phi^a$. The corresponding
algebra is the Weil algebra of $\g$, whic is defined as the tensor
product of
the exterior and symmetric algebras on $\g^*$,
$$
W(\g) = S(\g^*) \tensor \Omega(\g^*)
$$
where elements of $\Omega(\g^*)$ are
multilinear antisymmetric forms on $\g$, generated by anticommuting
basis of one-forms $\eta^a ~(a=1 \ldots m)$.  We introduce the
grading
one to $\eta^a$, and grading two to the commuting basis elements
$\phi^a$ of $S(\g^*)$.

The Weil model of equivariant cohomology is based on the Weil
algebra. Due to the similarities of the Weil and the BRST models, we
will handle them together, later on. Here we only discuss the Weil
algebra.

In the following we shall construct the differential calculus in a
Hamiltonian manner. For this we realize derivatives and internal
multiplications using Poisson brackets. For example contraction of
the
one-forms $\eta^a$ is realized by anticommuting ${\cal P}_a$ and
derivation with respect to $\psi^a$ is realized by commuting $\pi_a$
using Poisson brackets
\[
\poisson{\pi_a}{\phi^b} ~=~
\poisson{\p_a}{\eta^b} ~=~ \delta_a^b  \.
\]
In terms of these variables, the coadjoint action of $G$ on $W(\g)$
is
generated by the derivations
\be
L_a = - \f{abc}(\phi^b \pi_c + \eta^b \p_c)  \ ,   \mlabel{weillie}
\ee
the action being
$$
\poisson{L_a}{ \phi^b} = \f{abc} \phi^c \ , \ \ \ \ \ \ \
\poisson{ L_a}{ \eta^b} = \f{abc} \eta^c
$$
and
$$
\poisson{L_a}{L_b} = \f{abc} L_c \  .
$$

Next we introduce a couple of differential operators  on
$W(\g)$. We first define the "abelian" differential
\be
d_o ~=~ \phi^a \p_a \
\mlabel{abeldiff}
\ee
which identifies $\phi^a$ as the differential of $\eta^a$.
The nonvanishing actions are
\be
d_o \  \eta^a ~=~ \phi^a \, \ \ \  \ \ \ \ \ \ \ \
d_o \ \pi_a ~=~ -\p_a \.
\mlabel{d0action}
\ee
The second operator we define is
\be
d_{\g} = -f^{abc} (\eta^a \phi^b \pi_c + \half \eta^a \eta^b \p_c)
\,
\mlabel{liecohomdiff}
\ee
which computes the $W(\g)$-valued Lie algebra cohomology of $\g$.
This is readily seen by re-writing {}\refe{liecohomdiff} in terms of
{}\refe{weillie}:
\be
d_{\g} ~=~ \eta^a L_a + \half \f{abc} \eta^a \eta^b \p_c \ ,
\mlabel{gbrs}
\ee
which is of the familiar form of a Lie algebra coboundary operator,
or a BRST operator  related to the constraints $\set{L_a}$
acting on $W(\g)$.

The sum of the derivations {\refe{abeldiff}} and
{\refe{liecohomdiff}}
is the {\it Weil differential}:
\be
d_w ~=~ d_o + d_{\g} \,  \mlabel{wdiff}
\ee
with the actions
\[
d_w \ \eta^a = \phi^a -\half \f{abc} \eta^b \eta^c \,
\]
\be
d_w \ \phi^a = - \f{abc} \eta^b \phi^c \.  \mlabel{dWaction}
\ee
These three operators are all nilpotent derivations of degree one
$$
d_w^2 ~=~ d_o^2 ~=~ d_{\g}^2 ~=~ 0
$$
and act as exterior derivations on $W(\g)$.

{}From \refe{dWaction} we conclude that the cohomology of
$d_w$ on $W(\g)$ is trivial.  Indeed, $d_w$ can be obtained
from $d_o$ by a canonical transformation which is of the
functional form
\be
Q ~\to~
e^{-\Phi} Q e^{\Phi} ~=~ Q + \{ Q ,
\Phi \} + \frac{1}{2} \{ \{ Q , \Phi \} , \Phi \} + \ldots
\mlabel{canonical}
\ee
with generating function
\be
\Phi ~=~ \Phi_1 ~=~ \half \f{abc} \eta^a \eta^b \pi_c  \mlabel{phi1}
\ .
\ee
That is,
$$
d_w ~=~ e^{-\Phi_1} d_o e^{\Phi_1} \.
$$
In particular, we conclude that the cohomology of $d_w$ must
also be trivial.

In {}\refe{dWaction} we immediately recognize the action of
an exterior derivative on a connection one-form $A \sim \eta^a$ and a
curvature two-form $F \sim \phi^a$ of a principal $G$-bundle, {\it
i.e.} the definitions of the curvature and the Bianchi identity,
\bastar
dA &=& F - \half \ \commut{A}{A} \ , \\
dF &=& - \ \commut{A}{F} \ .
\eastar
These relations explain the relevance of the Weil algebra as a
universal model of connections on $G$-bundles.  The connection and
curvature define a unique homomorphism from $W(\g)$ to the exterior
algebra $\Omega(P)$ over the bundle, known as the Weil homomorphism
which carries the algebraic connection and curvature ($\eta^a,
\phi^a$) to the geometric ones ($A,F$) {}\tcite{mq}{kanno}.
This is also why $W(\g)$ appears in equivariant cohomology theory: it
models the universal bundle isomorphically on the level of
cohomology.
The universal bundle, being a contractible space with free
$G$-action,
can be used to lift a non-free $G$-action on $M$ to a free $G$-action
on a related space with equivalent homotopy. This leads to the
topological definition of equivariant cohomology {\cite{ab}}.

\vali

Finally, we note that the action of $d_w$ on the contraction ${\p}_a$
yields the corresponding generator {}\refe{weillie} of the coadjoint
action:
\[
\poisson{d_w}{\p_a} ~\equiv~ d_w \p_a + \p_a d_w  ~\equiv~
\poisson{d_{\g}}{\p_a} ~=~  L_a \ .
\]
In particular, the derivation $L_a$ has the natural structure of a
Lie-derivative on $W(\g)$ that commutes with our differentials,
$$
\poisson{d_w}{L_a} ~=~ \poisson{d_{\g}}{L_a} ~=~ \poisson{d_o}{L_a}
=0 \ .
$$

\vali

{\it The BRST Model:} We are now in a position to define the BRST
model of equivariant cohomology. For this we consider the tensor
product $ W(\g)\tensor \Omega(M)$ of the Weil algebra with the
exterior algebra over $M$. In analogy with the canonical realization
of the Weil algebra, we realize derivation with respect to the
coordinates $z^k$ on $M$ canonically by $p_k$, represent the
one-forms
$dz^k$ by anticommuting variables $c^k$ and the contraction operating
on $c^k$ by $\bar c_k$. The pertinent Poisson brackets are
\be
\{ p_k , z^l \} ~=~
\{ \bar c _k , c^l \} ~=~ \delta_k^l
\mlabel{pbs}
\ee
We introduce a grading of the variables by defining $gr(\eta, \phi,
c,
z) = (1,2,1,0)$. In terms of (\ref{pbs}) the exterior derivative on
$\Omega(M)$ is
$$
d = c^k p_k
$$
and the contraction and Lie
derivative with respect to the vector fields ${\cal X}_a$ are
\[
i_a ~=~ {\cal X}_a^k \bar c_k
\]
\[
{\cal L}_a ~=~ {\cal X}_a^k p_{k} + c^k\partial_k {\cal X}_a^l \bar
c_l
\]
We recall the exterior derivative {\refe{abeldiff}} and define the
following exterior derivative on $W(\g) \tensor \Omega(M)$,
\be
s_o ~=~ d + d_o  \mlabel{abelbrst}
\ee
Since the cohomology of $d_o$ is trivial, we conclude that the
cohomology of $s_o$ on $W(\g) \tensor \Omega(M)$ equals the deRham
cohomology of $d$ on $\Omega(M)$.

By introducing the canonical conjugation {\refe{canonical}}, we
obtain
from {\refe{abelbrst}}
\be
s_w =  e^{-\Phi_1} s_o e^{\Phi_1} ~=~ d + d_w \, \mlabel{wmodel}
\ee
where $d_w$ is the Weil differential {\refe{wdiff}}.  This is the
differential of the Weil model of equivariant cohomology {\cite{ab}}.

If we introduce a further conjugation with
\be
\Phi_2 ~=~ - \eta^a i_{\X_a} \,
\mlabel{phi2}
\ee
we then find the following nilpotent graded derivation of degree
one on $W(\g) \tensor \Omega(M)$ {}\tcite{osvb}{kalkman},
\be
s ~=~ e^{-\Phi_2} (d+d_w) e^{\Phi_2} ~=~
d + d_w - \phi^a ~ i_a + \eta^a
\cL_a \, \mlabel{ecubrst}
\ee
which  gives the BRST model for equivariant
cohomology. It is the natural nilpotent extension of
{\refe{cartandiff}}, with the ghost version of the non-nilpotency of
the Cartan model {\refe{squs}}, and the Weil differential $d_W$
taking
care of nilpotency. By construction the cohomology of $s$ on $W(\g)
\tensor\Omega(M)$ equals the deRham cohomology of $d$ on $\Omega(M)$.

\vali

As shown in {}\cite{osvb}, \ by appropriately restricting
$s$ to a subcomplex of $W(\g) \tensor \Omega(M)$ we
obtain the $G$-equivariant cohomology of $M$. Indeed, if we
consider the $\eta^a$-independent (which restricts onto
$S(\g^*) \tensor \Omega(M)$) and $(\cL_a + L_a)\ $-invariant
(which picks up the $G$-invariant part) subcomplex, we
recover the algebra $\Omega_G(M)$ of the Cartan model
\refe{cartancplex}. Moreover, after this restriction the BRST
operator {\refe{ecubrst}} reduces to the Cartan model differential
{\refe{cartandiff}}.

Correspondingly, in the Weil model, after restricting to the {\it
basic} subcomplex, defined to be horizontal (annihilated by $\p_a +
i_a$) and $G$-invariant, the operator {\refe{wmodel}} describes
equivariant cohomology.

In order to properly restrict the domain of $s$, following
\cite{osvb} we introduce another nilpotent operator $\w$ such that
its kernel coincides with the desired $G$-invariant,
$\eta$-independent subcomplex. For this we introduce another
copy of the Weil algebra, $\bar W(\g)$. We denote the generators of
$\bar W(\g)$ by $\barphi^a$ and $\bareta^a$, they are the
$\g^*$-valued coefficients corresponding to $\eta^a$ independence
(generated by $\p_a$) and $G$-invariance (generated by $\cL_a +
L_a$),
respectively. Consequently the desired nilpotent operator $\w$ must
include the terms $$
\w ~=~ \bareta^a (\cL_a + L_a) -\barphi^a \p_a ~+~ ...
$$
In order to complete the construction of $\w$, we specify its action
on $\bar W(\g)$. Indeed, if we define this action to coincide with
the action
of the Lie algebra coboundary
operator $d_{\bar \g}$ {}\refe{liecohomdiff} on $\bar W(\g)$,
we find that the following operator
\be
\w ~=~   \bareta^a (\cL_a + L_a) -\barphi^a \p_a
\mlabel{tuplavee}
\ee
is nilpotent. If we also extend the action of $s$ to $\bar W(\g)$
by
\be
s ~=~ d + d_w + d_{\bar o} - \phi^a \ i_a + \eta^a \cL_a
\mlabel{fullbrst}
\ee
we then find that $s$ and $\w$ satisfy the nilpotent algebra
\be
\poisson{s}{s} = \poisson{\w}{s} = \poisson{\w}{\w} = 0 \,
\mlabel{swalg}
\ee
and the $G$-equivariant cohomology of $M$ is isomorphic to the
cohomology of $s$, restricted to the kernel of $\w$. This determines
the BRST model for the $G$-equivariant cohomology of $M$ which is
relevant for the construction of nonabelian generalizations
\cite{wtwodim} of the Duistermaat-Heckman integration formula
{\refe{dh}}. Loop space generalizations of the constructions above
can
be found in {\cite{alushta}}.

The restriction onto the basic subcomplex in the Weil model can be
formulated using a nilpotent operator as well. The natural choice is
\be
\w_w=d_{\bar\g}+\bareta^a(\cL_a+L_a)-\barphi^a(\p_a + i_a) \,
\mlabel{wmw}
\ee
with the corresponding extension of {\refe{wmodel}}
\be
s_w=d+d_w+d_{\bar o} \.
\mlabel{fwm}
\ee
Operators {\refe{wmw}} and {\refe{fwm}} are canonical
transformations of {\refe{tuplavee}} and {\refe{fullbrst}},
respectively, with the generating function $-\Phi_2$. Thus they obey
an algebra similar to {\refe{swalg}}.
\vskip 1.0cm

\noindent
{\bf 4. Non-abelian equivariant symplectic two-forms}
\vali

In {\refe{equivham}} we presented the abelian equivariant extension
of
the symplectic two-form $\omega$. We shall now construct the most
general non-abelian equivariant extension of $\omega$ on the complex
$W(\g) \tensor \Omega(M)$, corresponding to the symplectic action of
the non-abelian Lie-group $G$ on the symplectic manifold $M$. In
analogy with {\refe{dh}}, this non-abelian equivariant extension can
then be used as the starting point for constructing non-abelian
generalizations of the DH integral that can be evaluated using
(non-abelian) localization methods.

We consider the Poisson bracket realization of $G$,
\be
\{ H_a , H_b \} ~=~ f^{abc} H_c  \, \mlabel{ahh}
\ee
with $H_a$ functions defined on the manifold $M$. In order to
construct the most general non-abelian generalization of
{\refe{equivham}}, we then introduce the following Ansatz,
\be
\h_o = \omega + \alpha \eta^a d H_a
+ \beta \phi^a H_a + \gamma \f{abc}\eta^a H_b \ d H_c \,
\mlabel{plaah}
\ee
which is the most general form of degree two that can be
constructed on $W(\g) \tensor \Omega(M)$ in terms of the variables
that appear in the BRST model of the $G$-equivariant cohomology. We
shall now determine the parameters in {\refe{plaah}} by requiring
$G$-equivariance, {\it i.e.} that $\h_o$ is annihilated both by
$s$ and $\w$ of the BRST model.

By demanding
\be
s_o \ \h = 0 \mlabel{poh}
\ee
we first get the conditions
\be
\alpha = - \beta~, ~~~~~~~~ \gamma=0~ \. \mlabel{conds}
\ee
We then introduce the canonical transformation
(\ref{canonical}) generated by
\be
\Phi_T ~=~ \Phi_1 + \Phi_2 \, \mlabel{phit}
\ee
with $\Phi_1$ and $\Phi_2$ defined in (\ref{phi1}, \ref{phi2}).
This yields for  $\h_o$,
\[
\h_o ~\to~ \exp\{ - \Phi_T\} \h_o \exp\{ \Phi_T\}
{}~=~ \h
\]
Explicitly,
\[
\h = \omega - \alpha \phi^a
H_a + (\alpha -1)\eta^a dH_a + \half (1 - \alpha) \f{abc} \eta^a
\eta^b H_c 
\]
As a consequence of {\refe{poh}}, $\h$ satisfies the condition $s\h =
0$.  In order to restrict it to the subcomplex $\Omega_G(M)$ we then
require that $$
\w \ \h ~=~ 0
$$
which sets
$$
\alpha \ = \ 1
$$
and yields
\be
\h ~=~ \omega ~-~ \phi^a H_a \mlabel{rsk}
\ee
as the most general $G$-equivariant extension of the
symplectic two-form $\omega$. We note that the final result
{\refe{rsk}} coincides with the nonabelian equivariant extension of
$\omega$ introduced in {}\cite{wtwodim}.

\vskip 0.3cm

In a number of applications to two-dimensional integrable models
(most
notably the KdV model) we obtain the following generalization:
instead
of {\refe{ahh}}, the Hamiltonians $H_a$ obey the centrally extended
Lie algebra
\[
\{ H_a , H_b \} ~=~ f^{abc} H_c ~+~ \kappa_{ab} \,
\]
with $\kappa_{ab}$ \ the Lie-algebra two-cocycle. Now the most
general
Ansatz of degree two for the equivariant extension of $\omega$ is
\[
\h_o = \omega + \alpha \eta^a d H_a + \beta
\phi^a H_a + \gamma \f{abc}\eta^a H_b \ d H_c
+ \mu \kappa_{ab} \eta^a \eta^b
\]
Requiring
\[
s_o \h_o ~=~ 0
\]
we then find, in addition to {\refe{conds}}, the condition
\[
\mu ~=~ 0 \.
\]
Performing the canonical transformation (\ref{canonical}, \ref{phit})
we get
\[
\h = \omega - \alpha \phi^a H_a + (\alpha -1)\eta^a dH_a
+ \half (1 - \alpha) \f{abc} \eta^a \eta^b H_c + (\alpha -
\frac{1}{2})
\kappa_{ab} \eta^a \eta^b \.
\]
If we set $\alpha=1$ we get the following two-form,
\be
\h ~=~ \omega - \phi^a H_a + \frac{1}{2} \kappa_{ab} \eta^a
\eta^b \. \mlabel{hkapp}
\ee
However, if we operate on $\h$ by $\w$, we find
\[
\w (\kappa_{ab} \eta^a\eta^b) = (\barphi^a \eta^b - \phi^a\bareta^b
- \f{cda} \bareta^c\eta^d\eta^b) \kappa_{ab} \,
\]
so that the restriction to $\Omega_{G}(M)$ can not be implemented.
Indeed, we have recovered the fact \ab \ that equivariant extensions
of the symplectic two-form are in one-to-one correspondence to the
Poisson liftings of the symplectic action of the group.  However,
using (\ref{smallh}, \ref{kappeq}), we can write {\refe{hkapp}} in
the
trivially $s$-closed form
\[
\h ~=~ s (\vartheta + \eta^a h_a ) \,
\]
and we conclude that it is still possible to derive localization
formulas for Hamiltonians constructed from a central extension of a
non-abelian Lie algebra.

\vskip 1.0cm

\noindent
{\bf 5. Superspace Formulation}

\vali

In the previous sections we have developed the BRST picture of
equivariant cohomology in terms of the bosonic coordinates $z^k$ on
$M$ and fermionic variables $c^k \sim dz^k$, and two Weil algebras
over the Lie algebra $\g$ acting on $M$ generated by $\eta^a, \phi^a$
and $\bareta^a, \barphi^a$ respectively. Obviously the coordinates
$z^k$, $\phi^a$ and $\barphi^a$ can be interpreted as bosonic
coordinates in a superspace, with corresponding superpartners $c^k$,
$\eta^a$ and $\bareta^a$.  In order to represent the exterior algebra
on this superspace in a Hamiltonian framework, we introduce the
pertinent conjugate variables with the nonvanishing Poisson brackets
\be
\poisson{p_k}{z^l} = \poisson{\bar c_k}{c^l} =
\poisson{\pi_a}{\phi^b} = \poisson{\p_a}{\eta^b} =
\poisson{\barpi_a}{\barphi^b} = \poisson{\barp_a}{\bareta^b}
= \delta_a^b \ . \mlabel{coords}
\ee
The Hamiltonian realization of the abelian derivation on
$\Omega(M)\tensor W(\g) \tensor \bar W(\g)$ is then an extension of
{}\refe{abelbrst} that acts on $\bar W(\g)$ as well:
\be
s_o ~=~ d + d_o + d_{\bar 0} ~=~ c^k p_k + \phi^a \p_a + \barphi^a
\barp_a \.  \mlabel{fabelbrst}
\ee
The supercanonical transformation generated by {}\refe{phit} relates
\refe{fabelbrst} to the full BRST operator {}\refe{fullbrst}. By
performing  the
inverse canonical transformation generated by $-\Phi_T$ on the
restriction operator {}\refe{tuplavee}, we get a description of
equivariant
cohomology on the level of the abelian exterior derivative
\refe{fabelbrst}. In
particular, by construction this transformed restriction operator
\be
\w_o = \w + \phi^a (\f{abc} \eta^b \pi_c - i_a)
= d_{\bar \g} + \bareta^a (\cL_a + L_a) - \barphi^a (\p_a -
\f{abc} \eta^b \pi_c + i_a)  \mlabel{tuplaveenolla}
\ee
obeys
\[
\poisson{\w_o}{s_o} = \poisson{\w_o}{\w_o}= 0 \ .
\]

\vali

Now we want to combine the coordinates {}\refe{coords} into
superfields on some underlying superspace. For this, we introduce a
N=2 superspace with two grassmannian directions $\theta$ and
$\bartheta$, and define the superfields
\ba
\A^k &=& z^k + \theta c^k \cr
\E_k &=& \bar c_k \bartheta + \theta \bartheta p_k \cr
\A^{\ a}_\theta &=& \eta^a + \phi^a \theta \cr
\E_{\bartheta,a} &=& \bartheta \pi_a + \theta \bartheta \p_a
\mlabel{sufield} \\
\A^{\ a}_\bartheta &=& - \bareta^a - \barphi^a \theta \cr
\E_{\theta,a} &=& \bartheta \barpi_a + \theta \bartheta \barp_a
\nonumber
\ea
The fields $\{\A^k\}$ generate the exterior algebra $\Omega(M)$, the
$\theta$-component $\A_\theta$ generate the Weil algebra $W(\g)$ and
the $\bartheta$-component the extra Weil algebra $\bar W(\g)$.

Notice that these superfields are truncated: In order to get the
full,
untruncated superfields it is necessary to double the number of
component fields. In the case of BRST quantization of a constrained
system these extra fields would be related to a BRST gauge fixing of
the theory, but in the following such fields are not relevant. We
refer to {\cite{mn}} and {\cite{hnt}}, where this aspect has been
discussed in the context of topological Yang-Mills theory.

{}From (\ref{coords}) we conclude, that the superfields
{\refe{sufield}} satisfy the (properly truncated) superspace Poisson
brackets
\[
\poisson{\E_{\alpha,a}(\zeta)}{\A^{\ b}_\beta(\zeta')} = g_{\alpha
\beta} \ \delta_a^b \ \delta(\zeta - \zeta ') \,
\]
where $\zeta$ denotes $\theta$ and $\bar \theta$ collectively and
$\alpha$ labels the components of $\E$ and $\A$ in {\refe{sufield}},
{\it e.g.} $\E_\alpha$ has components $(\E_a,
\E_\theta, \E_\bartheta)$, and the metric in $\theta$-space
is antisymmetric: $g^{\theta \bartheta} = - g^{\bartheta\theta} =
-g_{\theta\bartheta} = g_{\bartheta \theta} = 1$. The
$\delta$-function in the anticommuting variables is the appropriate
truncation of
\[
\delta(\zeta - \zeta ') = \theta \bartheta - \theta' \bartheta -
\theta \bartheta' + \theta' \bartheta' \,
\]
corresponding to our truncation of the superfields {}\refe{sufield}.

\vskip 0.2cm
We define the (truncated) supergenerators of infinitesimal
$G$-transformations
\[
\D_a^k ~=~ - \X_a^k + (d \X_a^k) \ \theta ~=~ - \X_a^k + c^l (\del_l
\X_a^k) \ \theta
\]
and the corresponding generators
\be
\G_a ~=~ \D_a^k \E_k ~=~ \bartheta  \X_a^k \bar c_k - d\X_a^k \bar
c_k
\theta \bartheta ~=~ \bartheta i_a  - \theta \bartheta \cL_a
\mlabel{sgauss}
\ee
that satisfy the Lie-algebra  {\refe{ahh}} in the superspace,
\be
\poisson{\G_a(\zeta)}{\G_b(\zeta')} = -\f{abc} \G_c(\zeta)
\delta(\zeta-\zeta') \. \mlabel{galg}
\ee
On the superfields,  {\refe{sgauss}}  generate the superspace gauge
transformations,
\ba
\poisson{\G_a(\zeta)}{\A^k(\zeta')} &=& \D_a^{k}(\zeta')
\delta(\zeta - \zeta')\cr
\poisson{\G_a(\zeta)}{\E_k(\zeta')} &=& - (\del_k \G_a(\zeta'))
\delta(\zeta - \zeta') \nonumber \,
\ea
where the truncated $\delta$-functions  give {\it
e.g.} $\D_a^{k}(\zeta')$ after integrating over the supercoordinates
$\zeta$.

We also introduce covariant derivation and components of the gauge
generators
in the $\theta$-direction:
\bastar
\D_{\theta,a}^{\ \ \ b} &=& \delta_a^{\ b}\del_\theta +
\f{abc}\A_\theta^c \\
\G_{\theta,a} &=&  g^{\theta \bartheta} \D_{\theta,a}^{\ \ \
b}\E_{\bartheta,b}
\,
\eastar
and similarly for $\bartheta$.

\vskip 0.3cm

We are now in a position to introduce the superfield representations
of the various quantities we have introduced previously. Indeed, we
find that the generator of coadjoint action on the Weil algebra
\refe{weillie} and the Lie algebra cohomology differential
\refe{liecohomdiff} have the following representations in terms of
the superfields:
\ba
L_a &=& - \int d\bartheta d\theta \f{abc }\A_\theta^{\ b}
\E_{\bartheta,c} \cr
d_\g &=& - \half \int d\bartheta d\theta \f{abc} \A_\theta^{\ a}
\A_\theta^{\
b} \E_{\bartheta,c} \mlabel{sliecd}
\ea
In addition we note that
\be
\A_\theta^{\ a} \G_a =  \phi^a i_a - \eta^a\cL_a \, \mlabel{aagee}
\ee
and that the abelian differential {}\refe{fabelbrst} can be expressed
as the generator of $\theta$-trans\-lations:
\be
s_o = g^{\alpha\beta}(\del_\theta \A_\alpha^{\ a}) \E_{\beta, a} \.
\mlabel{sabelbrst}
\ee
(Here we use the convention that a summation over $a$ is understood
only if the corresponding fields carry a representation of $G$ -
generically in
 the $\theta$ and $\bartheta$ components. An integration over
$d\bartheta  d
\theta$ is also understood here and in the following whenever it is
plausible.)

Combining these, we finally get the following superfield
representation of the
BRST operator {}\refe{fullbrst} for the equivariant cohomology:
\be
s = g^{\alpha\beta}(\del_\theta \A_\alpha^{\ a}) \E_{\beta, a} +
\A_\theta^{\
a} \G_a -  \half \f{abc} \A_\theta^{\ a} \A_\theta^{\ b}
\E_{\bartheta,c}
\.
\mlabel{supers}
\ee
In particular, in the last two terms we recognize the functional form
{\refe{gbrs}} of a BRST operator related to the constraint algebra
$\poisson{\G_a}{\G_b}=\f{abc}\G_c$, with $\A_\theta$ viewed as the
ghost field.

In order to obtain a superspace representation of {}\refe{tuplavee}
it
is easiest to work in terms of the canonically transformed
\refe{tuplaveenolla}.  Using (\ref{sliecd},\ref{aagee}) we then get
\be
\w_o = \A_\bartheta^{\ a}(\G + \G_{\theta})_a - \half \f{abc}
\A_\bartheta^{\
a} \A_\bartheta^{\ b}
\E_{\theta,c} \.
\mlabel{suplaveenolla}
\ee

Hence we conclude that superspace functions invariant under
$\theta$-translations describe $G$-equivariant cohomology on $M$,
provided that we restrict them to the kernel of $\w_o$.

Notice that the superspace representation {}\refe{suplaveenolla} has
the functional form of a conventional BRST operator related to the
constraints $(\G+\G_{\theta})_a$ generating the action of $G$ on
$\Omega(M)\tensor W(\g)$, {\it i.e.} on the $\A^k$ and $\A_\theta$
sectors of superspace. The corresponding ghosts $\A_\bartheta$
generate the extra Weil algebra $\bar W(\g)$. In particular, the
reducibility of the BRST operator {}\refe{fullbrst} is in some sense
lifted when the theory is formulated in the superspace.

\vskip 1.0cm

\noindent
{\bf 6. Relation to first stage reducible constraints}

\vali

We observe that the superfield formalism we have developed here is
identical to the superfield formulation of four dimensional
topological Yang-Mills theory developed in {\tcite{mn}{hnt}}; the
only
difference is that the dependence on space coordinates $\bf \vec x$
has been truncated. In the present section we shall discuss this
connection, and in particular how the BRST model for equivariant
cohomology is related to the Hamiltonian approach to constrained
quantization developed by Batalin, Fradkin and Vilkovisky (BFV) \bfv.

\vskip 0.2cm
In the BFV approach to Hamiltonian BRST quantization of constrained
systems, the BRST operator appears as a nilpotent operator that
encompasses all information about the algebra of constraints. For
example, nilpotency of {}\refe{suplaveenolla} in the framework of a
constrained system would immediately tell us that the constraints
$(\G
+ \G_{\theta})_a$ satisfy the algebra {}\refe{galg}.

In the case of a first class, first stage reducible constrained
system
we are dealing with constraints $F_a=0, a=1...m$ with Poisson
brackets
that close with some structure functions ${\cal C}^{abc}$,
\[
\poisson{F_a}{F_b} ~=~ {\cal C}^{abc}F_c
\]
and reducibility implies that there exists
$k\leq m$ linear relations between the constraints of the form
\[
B_i^a F_a ~=~ 0
\]
with some multipliers $B_i^a$. In the BFV approach we attach to the
constraints ghost fields $c^a$ together with their canonical
conjugates $\bar c_a$, and interpret the reducibility condition as an
extra constraint acting on the conjugate ghosts,
\[
B_i^a \bar c_a = 0 \. 
\]
which gives rise to ghost for ghost fields $\phi^i$ that are bosonic
fields with ghost number ({\it i.e.} grading) 2.  This reducibility
is
then incorporated into the nilpotency of a BRST operator in a
systematic manner, as explained in {\cite{bfv}}.

Here it is sufficient to consider - in analogy with four dimensional
toplogical Yang-Mills theory {\tcite{wtym}{bbrt}} - a constraint
algebra that consists of two intrinsically irreducible sets, a set
$\{E_k\}$ of abelian constraint functionals and a (not bigger) set of
constraint functionals $\{G_a\}$ which generate a non-abelian Lie
algebra with structure constants $\f{abc}$.  These two sets of
constraint functionals then form a reducible constraint algebra with
multipliers $\delta_a^b$ and $-D^k_a$ respectively, obtained by
setting
\be
G_a - D^k_a E_k  ~=~ 0 ~~~ \ \forall a \. \mlabel{redu}
\ee
so that the structure functions of the constraint algebra are
\bastar
\C^{abc} &=& \f{abc} \cr
\C^{abk} &=&0 \\
\C^{akl} &=& \poisson{D_a^l}{E_k}
\eastar
This is exactly the case which leads to equivariant cohomology: The
abelian constraint functionals $E_k $ can be identified as conjugate
momenta of local coordinates $z_k$ on some manifold, and the
constraints $E_k =0$ imply independence of the coordinates, \ie that
we are interested in cohomological properties of the corresponding
manifold.  The ghosts of the topological constraints constitute a
basis for one-forms on the manifold.  The nonabelian constraints
generate the group $G$ acting on the manifold, and the cohomology of
the abelian BRST operator (exterior derivative) $$ s=c^a E_a $$
reduces to the $G$-equivariant cohomology. The ghosts of the
nonabelian constraints are the $\eta^a$ fields, which generate the
Weil algebra together with the ghosts for ghosts $\phi^a$.  Finally,
equation {}\refe{redu} is just the generating vector field
\refe{vcomm} written in component form.

In particular, the formalism presented here is identical to that
found
in the Hamiltonian quantization of the four dimensional topological
Yang-Mills theory {}\tcite{mn}{hnt}, where the reducibility equation
\refe{redu} is the Gauss law constraint.

In the case of group actions on finite dimensional spaces we already
have a BRST operator {}\refe{ecubrst} which encorporates all
information about the constraint algebra $\set{E_k, G_a}$.  In
particular, the correspondence between the constraint algebra and
exterior calculus is
\bastar
d &=& c^a E_a \cr
i_a &=& D_a^k \bar c_k  \cr
\cL_a &=& G_a + c^k\  \C^{alk} \ \bar c_l
\eastar
Writing {}\refe{ecubrst} in terms of these, we get
\be
s = c^a E_a + \eta^a G_a + \phi^a (\p_a - D_a^k \bar c_k)
-\half \f{abc} \eta^a \eta^b \p_c - \C^{akl} \eta^a c^k \bar c_l
- \f{abc} \eta^a \phi^b \pi_c \.
\mlabel{minimal}
\ee
where we recognize terms corresponding both to the two sets of
original constraints and to the ghost constraint, terms related to
the
structure functions $\C^{abc}$, and an additional term related to the
Lie-algebra cohomology operator {}\refe{liecohomdiff}.

As shown by  (\ref{wmodel}, \ref{ecubrst}) the cohomology captured by
\refe{minimal} is just the deRham cohomology of the abelian BRST
operator $d=c^a E_a$. To get a nontrivial answer, a restriction onto
the basic subcomplex should be made,  using $\w$. This is an easy
way to establish the nontriviality of observables in topological
Yang-Mills.

As discussed in {}\cite{hnt}, the minimal BRST operator
{}\refe{minimal} can be extended by adding more fields. For gauge
fixing purposes (and to maintain manifest Lorentz invariance in field
theory applications), the Lagrange multipliers corresponding to the
constraints should be made dynamical. Doing this, extra abelian
constraints arise, which express that the multiplier momenta must
vanish.  The corresponding ghost fields are the so called
anti-ghosts. In this way the fields related to the constrained system
$\set{E_k=0, G_a=0}$ fall naturally in three sets \hnt:

- The first set includes the fields related to the topological
constraint $E_k=0$: the coordinate $z$, the ghost $c$, an antighost
and a multiplier, as well as the corresponding momenta.

- The second set includes the Weil algebra generated by the ghost for
ghost $\phi$ and the ghost $\eta$ of the constraint $G_a=0$, and an
antighost and a multiplier for the same.

- The third set includes an antighost and a multiplier for the ghost
constraint. Again, to maintain manifest Lorentz invariance in field
theory applications, some extra fields have to be defined. These are
the extra ghosts of \bfv, a commuting and an anticommuting canonical
pair for each reducibility equation, \ie for each $G_a$. We recognize
in the extraghost the generators $\barphi,\bareta$ of our extra Weil
algebra $\bar W(\g)$.

The fields in each of these three sets define a component of
the superconnection $\set{\A_k, \A_\theta, \A_\bartheta}$ in
{}\refe{sufield} in the absence of all multiplier and
antighost fields, in analogy {}\tcite{mn}{hnt} with four
dimensional topological Yang-Mills theory. In particular,
{}\refe{supers} reproduces the BRST operator of topological
Yang-Mills.

Analogues to conjugations (\ref{phi1}, \ref{phi2}) can also
be found in {}\tcite{mn}{hnt}. In {}\cite{hnt} other conjugations
are introduced as well. The most interesting one is
generated by
\[
\Phi_3 = \f{abc} \eta^a \bareta^b \bar\pi_c \,
\]
which lifts the abelian action of $d_{\bar 0}$ on $\bar
W(\g)$ to the coadjoint action:
\[
s \to s' ~=~ s + \eta^a \bar L_a + \f{abc} \phi^a \bareta^b \barpi_c
\equiv s + \A^a_\theta \D^{\ ab}_\bartheta \E_\theta^b \.
\]
 On the superspace level this BRST operator is related to
the {\it full} superspace Gauss law, extended to act on
$\A_\bartheta,\E_\theta$ as well:
\[
\poisson{\E_{\theta,a}}{s'} = g^{\alpha\beta}\D_{\alpha,a}^{\ \ \ b}
\E_{\beta,b} \equiv (\G+\G_\theta+\G_\bartheta)_a \.
\]
Notice however, that in $s'$ the roles of
$\A_\bartheta,\E_\theta$ as superspace ghosts as in
{}\refe{suplaveenolla} has been lost.

\vskip 1.0cm

{\bf 7. Conclusions}

\vali

Following \osvb \ we have formulated equivariant cohomology
in the context of localization formulas in terms of two
nilpotent operators, the BRST operator $s$ and the
restriction operator $\w$. In addition, we have developed a
superfield formalism for equivariant BRST using a N=2
superspace with fermionic coordinates $\theta,\bartheta$. We
have found, that in this superspace formalism all variables
relevant to localization can be combined into a single
superconnection $\A$.

Furthermore, we have shown that the BRST operator can be
conjugated to the translation operator in the
$\theta$-direction in superspace, and the restriction
operator acquires the form {}\refe{suplaveenolla} of a
conventional BRST operator related to the {\it superspace}
action of the $\A_k$ and $ \A_\theta$ parts of the {\it
superspace} Gauss law, with the remaining superfields
$\A_\bartheta$ acting as superghosts of this
superconstraint.  The connection of equivariant cohomology
and BFV quantization of four dimensional topological
Yang-Mills theory becomes then transparent.

Depending of the interpretation of the fields $\A_k$, the
superfield formalism presented here describes both
equivariant cohomology in the symplectic setting relevant to
localization, and the BRST structure of (cohomological)
topological field theories. From this we conclude that there
should be a unified description of localization in the
symplectic loop space {}\cite{omat}, the supersymmetric loop
space {}\cite{palo} and in the case of topological field
theory {}\cite{wtym}. Inded, this is consistent with the
mathematical conjecture {}\cite{ward} that all lower
dimensional integrable models could be obtained as
dimensional reductions of 4-dimensional self-dual Yang-Mills
theory, which is intimately connected with topological
Yang-Mills.

\vali

\vali

\noindent
{\bf Acknowledgements:} A.N. acknowledges a discussion with
R. Stora that prompted the present investigation.

\pagebreak

\end{document}